\documentclass[]{piparticle-final}
\usepackage{graphicx}
\usepackage{amsmath}

\usepackage{cite} % To compress citation ranges
\usepackage{epstopdf} 

\begin{document}

\volume{7}               % To be inserted by Editor
\articlenumber{070004}   % To be inserted by Editor
\journalyear{2015}       % To be inserted by Editor
\editor{L. A. Pugnaloni}   % To be inserted by Editor
\reviewers{G. Lumay, University of Li\`ege, Belgium.}  % To be inserted by Editor
\received{17 November 2014}     % To be inserted by Editor
\accepted{2 March 2015}   % To be inserted by Editor
\runningauthor{P. Mort}  % To be inserted by Editor
\doi{070004}         % To be inserted by Editor

\title{Characterizing flowability of granular materials by onset of jamming in orifice flows}

% Institution references with \cite are inserted after \maketitle in theaffiliation enviroment
\author{Paul Mort\cite{inst1}\thanks{Email: mort.pr@pg.com}
	}

\pipabstract{
This paper describes methods to measure flow rate and jamming onset of granules discharged through a flat-bottom cylindrical hopper with a circular orifice. The intrinsic jamming onset for ideal particles (spherical, monodisperse, smooth) is experimentally measured by two independent methods, with good agreement. For non-ideal particles, the normalized jamming onset increases with elongated granule shape, broadened size distribution and increased friction as measured by the drained angle of repose. An empirical model of the jamming onset is introduced to quantify these effects over the range of materials investigated.  The jamming onset can be used as a measure of differentiation between relatively free-flowing granules.}

\maketitle

\blfootnote{
\begin{theaffiliation}{99}
	\institution{inst1} Procter \& Gamble Co., 5280 Vine Street, Cincinnati, OH 45217, USA.
\end{theaffiliation}
}

%\maketitle

\section{Background}

The motivation for this work is to find a means of characterizing and differentiating flow quality of relatively free-flowing granular materials.  On the one hand, industry requires methods that can be performed relatively simply and reproducibly.  On the other hand, both industry and academia seek better fundamental understanding of granular rheology via physical mechanisms governing flows; in the case of industry, fundamental understanding should extend to commercially-relevant granular materials.  This paper correlates the onset of jamming with granular characteristics, including particle size distribution and particle shape.  Details are provided on size and shape characterization of commercially-relevant materials. 

Characterization of cohesive powder flow has been relatively well established using shear cells to quantify incipient bulk flow (i.e., yield loci) of powders and granules.  Flow functions calculated using yield loci can differentiate between strongly-cohesive, mildly-cohesive and free-flowing materials \cite{i}; this is highly relevant to bulk storage and handling of powders.  However, shear cell measurements are relatively insensitive in regards to differences among freely-flowing granules.  For applications where the quality of granular dynamic flow is of interest, we need a more sensitive methodology, hence the motivation of the current work. Various approaches are discussed in the literature, including using rotating drums \cite{ii,iii} and impeller-driven flows \cite{iv,v} to measure granular rheology.  The current work focuses on the onset of jamming in orifice flows to differentiate the quality of otherwise freely-flowing granules.

Let us consider jamming in the context of a shear cell analysis.  According to continuum modeling, a ``cohesionless'' material measured using a shear cell (i.e., a yield locus passing through the origin) implies flow through a hopper cone opening of infinitesimal diameter. At this limit, the continuum model fails because the orifice must be at least as large as the particle diameter.  Further, local packing and jamming effects require the opening to be significantly larger than a single particle \cite{vi}.  A theoretical analysis of intrinsic cohesion suggests a boundary layer effect of 3-5 particle diameters \cite{vii}; as a starting point, this can be taken as a theoretical minimum dimensionless orifice size for an otherwise free-flowing granular material.  Experimental studies on the physics of jamming in granular flows report jamming transitions on the basis of a dimensionless hopper opening at the onset of the jamming transition \cite{viii,ix}, where the dimensionless 
hopper opening ($D_o/d$) is defined by the ratio of the opening size ($D_o$) relative to a monodisperse particle size ($d$).  The current work is an empirical investigation, seeking to elucidate the effects of both size distribution and shape factors on the jamming probability of free-flowing granules. 

\section{Experimental}

\begin{figure}
\centering
 \includegraphics[width=0.5\columnwidth]{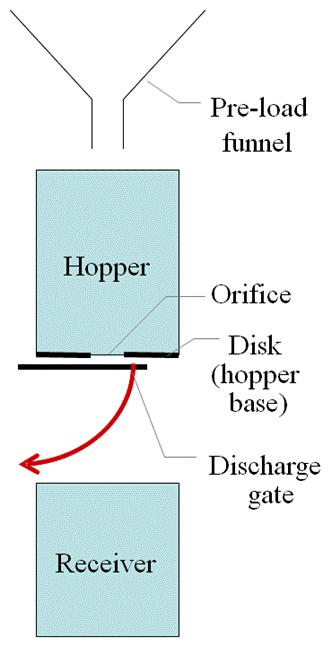}
 \caption{Schematic of a flat-bottom cylindrical hopper device (Flodex\texttrademark). }
 \label{fig1}
\end{figure}

\begin{figure}
\centering
 \includegraphics[width=0.8\columnwidth]{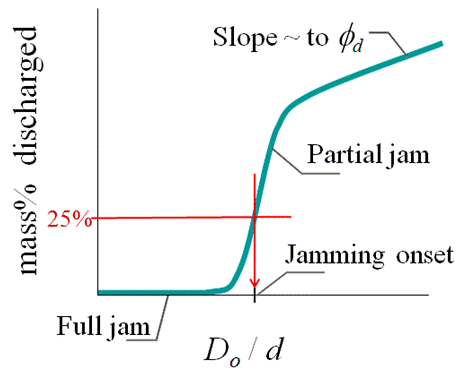}
 \caption{Interpolation of Jamming Onset ($J$) at 25\% mass discharged.}
 \label{fig2}
\end{figure}

The experimental device used in this work was a flat-bottom cylindrical hopper with a set of interchangeable disks having a range of discrete orifice sizes; it is available as a commercial flow testing system, Flodex\texttrademark (Hanson Research, Chatsworth, CA, USA), with additional disks machined as required to extend the range of orifice diameters.  The range of orifice sizes used with the Flodex included:  2.0, 2.5, 3.0, 3.5, 4, 5, 6, 7, 8, 9, 10, 12, 14, 16, 18 and 20 mm.  Experiments were done with common materials including narrowly-classified glass beads, Ottawa sand, and a variety of granular detergent samples.  In all cases, these samples were relatively free-flowing according to shear cell measurements.

A schematic of the orifice flow instrument is shown in Fig.~1.  Approximately 100 ml of material is used in the test, filling the 5.7 cm diameter stainless-steel hopper to a height of about 4 cm by pouring the granular sample through the loading funnel.  After the sample settles, the spring-loaded discharge gate is opened and the sample is allowed to drain through the orifice into a receiving cup below.  Once the flow stops and remains stopped for 30 seconds, the mass of discharged material is measured.  Clogging is defined as a persistent jammed state where the orifice remains obscured by the granules at the point of flow stoppage.  For each measurement, the mass\% discharged is calculated according to the formula:  (mass\% discharged) = 100 $\times$ (mass discharged) / (sample mass).  The average of the three mass\% discharge calculations is plotted as a function of the dimensionless orifice size ($D_o/d$), with the mass\% discharged on the ordinate and the dimensionless orifice size on the abscissa.  This 
procedure is repeated using incrementally larger orifice sizes until the hopper discharges without clogging for three consecutive trials.  The averaged data are linearly interpolated to find the Jamming Onset ($J$), defined here as the value of the dimensionless orifice size at the point of 25 mass\% average discharge (Fig.~2). 

For trials that drain without clogging, the drained angle of repose, $\phi_d$, is calculated according to Eq. (1), derived assuming the granular material discharges in the form of an inverted cone with a base diameter equal to the hopper diameter, $D_h$, and truncated where the cone’s apex protrudes through the orifice with diameter, $D_o$.  Assuming the remaining volume of retained material has cylindrical symmetry (Fig.~3), simple solid geometry and the material's repour (loose) bulk density, $\rho_{bulk}$, are sufficient to relate the measured mass of retained material, $M_{ret}$, to the drained angle of repose.

\begin{equation}
\phi_{d}=\arctan\left[ \frac{24 M_{ret}}{\pi \rho_{bulk} (2D_{h}^3-3D_{o}D_{h}^2+D_{o}^3)} \right] \label{1}
\end{equation}

Orifice size is made dimensionless by scaling the orifice size ($D_o$) to a characteristic particle size ($d$).  The Sauter mean ($d32$) is used as the characteristic size for this study.  Note the Sauter mean is weighted toward the surface area of the granules, the surface area being relevant to inter-particle contact and frictional interactions of the flow.  Studies of cohesive dry powders show correlation between the Sauter mean size and flow properties \cite{x}.  Size distributions were measured by sieving, fitting the size distribution using a log-normal distribution model to obtain a mass-based geometric mean ($d43$) and geometric standard deviation ($\sigma_g$) for each sample, then converting to the Sauter mean using the Hatch-Choate relation \cite{xi}, Eq. (2).  Examples of size distribution characteristics are shown in Fig.~4, for select samples A-D discussed in detail later in the results section.

\begin{figure}
\centering
 \includegraphics[width=0.6\columnwidth]{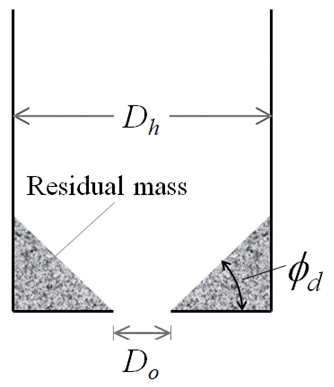}
 \caption{Vertical cross section of residual mass in a cylindrical hopper after draining successfully without a clog. }
 \label{fig3}
\end{figure}

\begin{figure}
\centering
 \includegraphics[width=0.98\columnwidth]{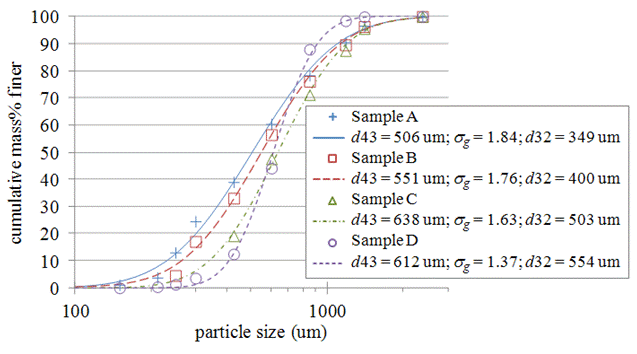}
 \caption{Examples of particle size distribution analysis for four samples, A-D.  The data points are from sieve analysis; the fitted curves are based on a log-normal distribution, weighted to the mass on each sieve. }
 \label{fig4}
\end{figure}

\begin{figure}
\centering
 \includegraphics[width=0.98\columnwidth]{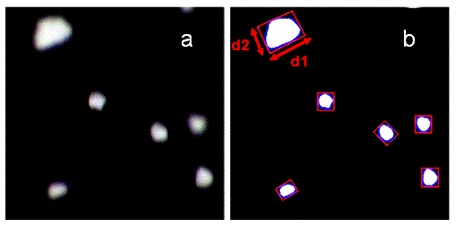}
 \caption{Image analysis freeze frame and cross-sectional aspect ratio (AR) analysis:  a) grayscale image; b) B\&W image resulting from threshold analysis with particles outlined in red, $AR = d1/d2$. }
 \label{fig5}
\end{figure}

\begin{figure}
\centering
 \includegraphics[width=0.8\columnwidth]{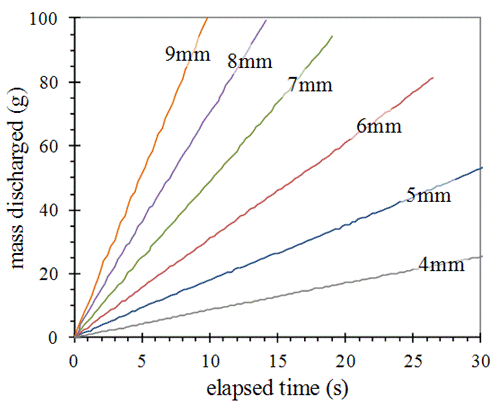}
 \caption{Discharge rate data for Sample D, measured over a series of orifice diameters (4 mm to 9 mm).  For each condition, the flow rate is the time derivative of mass discharged.  The flow rate data are plotted according to the Beverloo relation in Fig.~9. }
 \label{fig6}
\end{figure}

\begin{equation}
 \ln{d32}=\ln{d43}-\ln^2{\sigma_g}
\end{equation}

Shape characteristics were measured using an automated image analysis system, Solids Sizer (JM Canty, Buffalo, NY), using statistical averaging of $10^4$ cross-sectional particle images per sample.  An example of particles counted in a freeze-frame captured by the image analysis system is given in Fig.~5.  Note that the particles are in free-fall at the point of image capture, so the cross sectional image is randomly distributed over the possible states of rotation.  The median aspect ratio ($AR50$) was used as the characteristic shape factor in this study.  

The granular discharge rate was measured by placing the receiver cup on an electronic balance with data acquisition to record the discharge mass with time.  This experiment was done using orifice sizes above the jamming onset.  Discharge data are provided in Fig.~6, using Sample D as an example. Sample D is discussed in more detail in the results section.  The flow rate was determined by taking the time derivative (slope) of the cumulative discharge.  The mass flow rate is converted to a volumetric rate using the repour bulk density.

\section{Results and Discussion}

\subsection{Analysis of jamming onset using a multivariate model}

Results for nearly ``ideal'' materials including tightly classified glass beads and washed Ottawa sand are shown on Fig.~7.  These samples are nearly mono-disperse.  The glass beads are nearly spherical, while the sand is slightly irregular in shape.  Both have relatively smooth-surfaces and a low drained friction angle.  The jamming onset for the glass beads is in the range of 3 to 5 particle diameters.  The washed sand has a slightly higher jamming onset of about 7 diameters.  Other samples characterized in this work included commercial granules having similar compositions, but with a range of particle size and shape characteristics, the effect of which resulted in relative jamming onsets up to 30 and drained friction angles in excess of 60 Degrees.  Because the commercial samples have several aspects of variability, a multivariate model was formulated to assess the relative effects of friction, size distribution and shape characteristics.

\begin{figure}
\centering
 \includegraphics[width=0.98\columnwidth]{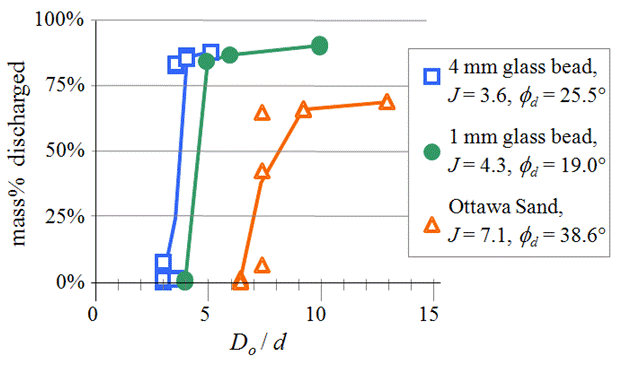}
 \caption{Cylindrical hopper discharge data, jamming onset calculations and drained angle of repose for glass bead and washed sand samples.}
 \label{fig7}
\end{figure}

\begin{figure}
\centering
 \includegraphics[width=0.98\columnwidth]{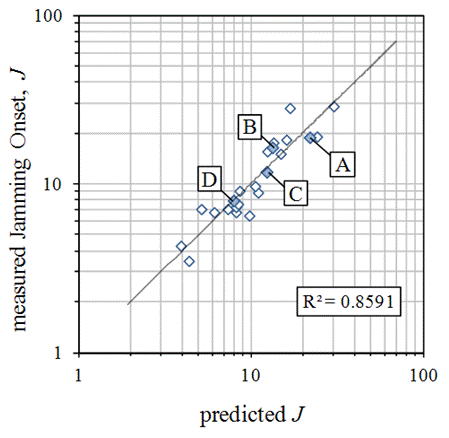}
 \caption{Multivariate regression of jamming onset model [Eq. (3)]; deviation from the diagonal represents uncertainty in the empirical model.  Solid data points labeled A-D indicate samples used in Flow Rate analysis (Fig.~9); their size distributions are shown in Fig.~4. }
 \label{fig8}
\end{figure}

\begin{figure}
\centering
 \includegraphics[width=0.98\columnwidth]{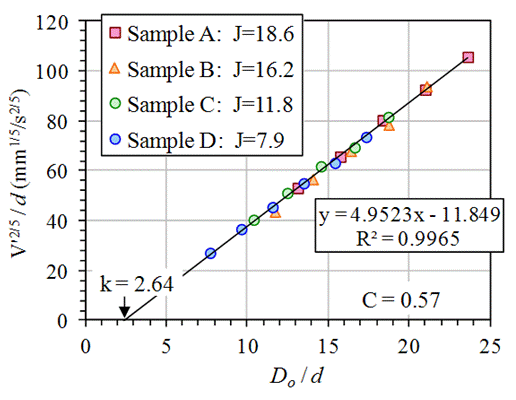}
 \caption{Regression [Eq. (5)] of flow rate data collected for selected samples (A-D) having a range of jamming onsets. }
 \label{fig9}
\end{figure}

Data from the broader set of 23 samples (Table 1), including commercial granular materials and manipulations thereof, were used to generate a multivariate power-law model for the jamming onset ($J$) as a function of particle parameters [Eq. (3)].  The model parameters include the shape factor (median aspect ratio, $AR50$), size distribution breadth (geometric standard deviation, $\sigma_g$) and an excess friction factor defined as $XF = \max [1, \tan(\phi_d)]$.  The range in breadth of size distribution was from about 1.0 to 2.2; the range of median aspect ratio was from about 1.0 to 1.4.  The fitted coefficients ($a$, $b$, $c$) represent exponents in the power-law form of the model.  While this equation is purely empirical, its form is logical in the sense that it reduces to an intrinsic jamming intercept ($k_j$) when the other terms are minimized.  In other words, extrapolation to $k_j$ represents the jamming onset of an idealized sample (mono-disperse, spherical, smooth, low 
friction). 

\begin{align}
 \ln(J)=&\ln(k_j)+a\ln(AR50) \notag\\ \label{3}
 &+b\ln(\sigma_g)+c\ln(XF)
\end{align}

The results of the multivariate model are shown graphically in Fig.~8.  While the correlation coefficient ($\sim$0.86) indicates that the model is reasonably predictive, the scatter suggests uncertainty, perhaps in the choice of model parameters, their interaction and/or uncertainty in measurements.  For example, the model may be over-simplifying the parameter space, ignoring potentially important factors such as more complex shape factors, the breadth of the shape distribution, and interactions between size and shape distributions. 

The statistical analysis is shown in Table 2.  The most statistically significant factors are the intrinsic jamming intercept, $k_j$, excess friction factor, $XF$, and the geometric size distribution, $\sigma_g$. The value of $k_j$ is $\sim$2.9, at the lower end of the range predicted by Wier \cite{vii}. While the particle shape factor, $AR50$, is somewhat less significant statistically, its high coefficient (2.38) indicates that particle shape can have a strong impact on jamming, even over the relatively narrow range tested. 

\begin{table}
 \caption{Experimental data for relative jamming onset ($J$) as a function of excess friction factor ($XF$), geometric size distribution ($\sigma_g$), and median aspect ratio ($AR50$).}
\begin{center}
\begin{tabular}{c r r r r}
\hline \hline
Sample & $J$ & $XF$ & $\sigma_g$ & $AR50$ \\
\hline
A&18.6&1.49&1.84&1.36 \\ 
B&16.2&1.01&1.76&1.35 \\
C&11.8&1.05&1.63&1.35 \\
D&7.9&1.00&1.37&1.26 \\
E&6.8&1.00&1.33&1.30 \\
F&7.1&1.00&1.34&1.24 \\
G&8.9&1.10&1.64&1.25 \\
H&9.6&1.08&1.48&1.32 \\
I&28.9&1.90&1.79&1.41 \\
J&15.1&1.00&2.14&1.26 \\
K&18.3&1.53&1.50&1.32 \\
L&28.2&1.24&1.79&1.34 \\
M&15.6&1.00&1.56&1.41 \\
N&19.1&1.83&1.56&1.41 \\
O&6.5&1.08&1.47&1.28 \\
P&7.3&1.00&1.39&1.27 \\
Q&9.1&1.05&1.41&1.26 \\
R&7.6&1.09&1.32&1.27 \\
S&17.6&1.73&1.31&1.26 \\
T&6.8&1.00&1.23&1.21 \\
U&7.1&1.00&1.10&1.20 \\
V&4.3&1.00&1.05&1.10 \\
W &3.5&1.00&1.05&1.15 \\
\hline \hline
\end{tabular}
\end{center}

\end{table}

\begin{table}
 \caption{Parameter estimates and standard errors obtained from multi-variate regression of Eq. (3) using the data of Table 1.  ``Prob$>|t|$'' is the probability that the true parameter value is zero, against the two-sided alternative that it is not; values less than about $0.05$ are typically regarded as highly significant.}
\begin{center}
\begin{tabular}{l r r r}
\hline \hline
 \small Term& \small Estimate& \small Std Error&\small P$>|t|$\\
\hline
$\ln(k_j)$&1.072&0.225&0.0001 \\
$a$&2.380&1.267&0.0758\\
$b$&1.404&0.398&0.0023\\
$c$&1.095&0.265&0.0006 \\
\hline \hline
\end{tabular}
\end{center}
\end{table}

\subsection{Flow rate measurements and Beverloo analysis}

Flow rate data were analyzed using the Beverloo equation \cite{xii}.  It is shown here in volumetric form [Eq. (4)], where $V'$ is the volumetric feed rate, $D_o$ is the orifice diameter, $d$ is the characteristic particle size (Sauter Mean), $g$ is the gravitational constant and $C$ and $k_f$ are fit parameters; $D_o - k_f d$ represents a reduced orifice size for active flow caused by a boundary layer that scales with doparticle size.  For the purpose of regression analysis as a function of relative orifice size ($D_o /d$), the Beverloo equation can be written in linearized form, Eq. (5), where $C$ is solved using the regression slope and $k_f$ using the intercept.

\begin{equation}
V'=C\sqrt{g} \left(D_o-k_f d\right)^{5/2} \label{4} 
\end{equation}

\begin{equation}
 \frac{V'^{2/5}}{d}=C^{2/5}g^{1/5}\left(\frac{D_o}{d}-k_f \right) \label{5}
\end{equation}

Several flow rate data sets are shown in Fig.~9.  These data sets (Samples A-D) represent similar materials, but with different shape and size distribution characteristics.  Sample A is a commercial granule; samples B and C are manipulated by classification; and sample D is classified then further rounded in a layering process.  The jamming onsets ($J$) of these samples span a significant range from about 8 to 20. However, the Beverloo regressions are remarkably similar.  All lie on a common slope ($C \sim 0.57$) and extrapolate to a similar zero-flow intercept ($k_f \sim 2.64$).   In other words, the granular flow behavior is consistent for orifice sizes above the jamming onset, but subtle differences in the granular characteristics and frictional properties have a significant effect on the onset of jamming.  This suggests that the onset of jamming is a useful measure to differentiate the flow behavior of relatively free-flowing granular materials.

Consider $k_f$ to represent an ``intrinsic jamming onset'' at which the flow rate goes to zero.  Note that the extrapolated value for intrinsic jamming (i.e., no flow) is only slightly lower than the power-law intercept of the multivariate model ($k_j \sim 2.9$).  Recall, the measurement of the jamming onset, $J$, is based on the interpolated orifice size for an average 25\% sample discharge from the test hopper.  Thus, it is logical that $k_j$, which is based on multivariate analysis of $J$, may slightly exceed $k_f$, which implies zero flow.  On the other hand, the slope of the jamming probability function tends to be quite steep in the range from zero to 25\% discharge.  So, for the purpose of this paper, we can merge the interpretation of $k_j$ and $k_f$ ---both represent an intrinsic jamming limit.  Indeed, even though each one was measured independently using different methods, their values are nearly identical.

An interpretation of this result is that the ``intrinsic jamming onset'' is characteristic of orifice flow for an ideal particle (monodisperse, spherical, no excess friction), and that differences between the intrinsic onset and the actual measured jamming onset are due, in large part, to irregularities in the size distribution, shape and frictional roughness of the granules. 

\section{Conclusions}

The statistical analysis of jamming is a promising approach for characterization of relatively free-flowing granular materials.  In this work, we evaluated the effect of dimensionless orifice size on the onset of jamming and flow rate.  A multivariate analysis of the jamming onset suggests an intrinsic jamming probability of about 2.9 particle diameters for an idealized sample; irregularities such as size distribution, shape or excess friction support jamming across an increased number of particle diameters.  While all three parameters are significant in the regression analysis, the result suggests that shape (aspect ratio) is an especially important contributor to jamming.  Analysis of the flow rate data using the Beverloo equation extrapolates to a similar value for the intrinsic jamming onset ($\sim$2.6 particle diameters), providing good experimental agreement between two independent methods.

The experimental methods described herein are relatively simple.  Continued work covering a wider range of materials can help to build stronger empirical models for the effects of shape, size distribution and other frictional properties on jamming probability.  In addition, first principle modeling of dense flows may help to elucidate the underlying mechanisms of jamming and the real effects of non-ideal particle characteristics, and perhaps help to build a more theoretical model.

Lastly, the reader should temper the results presented with the caveat that all of the samples used herein are relatively free-flowing granular materials.  As materials become more cohesive and/or compressible, it is not clear that the flow rate and jamming onset experiments will continue to converge on a common intrinsic jamming onset.

\begin{acknowledgements}
The original data for this paper were collected and analyzed as part of a student internship at Procter \& Gamble by Derek Geiger, University of Michigan, with the assistance of Mark Wandstrat at P\&G, and presented at the 2007 Annual Meeting of the American Institute of Chemical Engineers \cite{xiii}.  This work at P\&G followed an earlier collaboration with Professor R. P. Behringer \cite{xiv}; the author acknowledges Professor Behringer's guidance and insight on using fundamental granular physics to address problems of industrial relevance.
\end{acknowledgements}

\end{document}